\documentclass[aps,prl,twocolumn,a4paper,10pt,superscriptaddress]{revtex4-1}
\usepackage{amsmath}   
\usepackage{amssymb}

\usepackage{hyperref}  

\usepackage[english]{babel}
\usepackage{microtype}

\usepackage{url}
\usepackage{mathtools}
\usepackage{siunitx}
\usepackage{ifthen}
\usepackage{tikz}
\usetikzlibrary{arrows,positioning,backgrounds,calc}

\usepackage[utf8]{inputenc}
\DeclareUnicodeCharacter{014D}{\=o}

\graphicspath{{bilder/}}

\newcommand{\Na}{$^{22}$Na}
\long\def\/*#1*/{}

\usepackage{color}
\definecolor{gp1}{HTML}{1B9E77}
\definecolor{gp2}{HTML}{D95F02}

\begin{document}

\title{Spin-resolved Fermi surface of the localized ferromagnetic Heusler compound Cu$_{2}$MnAl measured with spin-polarized positron
annihilation}

\author{Josef A. Weber}
\email{josef-andreas.weber@frm2.tum.de}
\affiliation{Physik-Department, Technische Universit\"at M\"unchen, James-Franck Stra\ss e, 85748 Garching, Germany}

\author{Andreas Bauer}
\affiliation{Physik-Department, Technische Universit\"at M\"unchen, James-Franck Stra\ss e, 85748 Garching, Germany}

\author{Peter B\"oni}
\affiliation{Physik-Department, Technische Universit\"at M\"unchen, James-Franck Stra\ss e, 85748 Garching, Germany}

\author{Hubert Ceeh}
\affiliation{Physik-Department, Technische Universit\"at M\"unchen, James-Franck Stra\ss e, 85748 Garching, Germany}

\author{Stephen B. Dugdale}
\affiliation{H.H. Wills Physics Laboratory, University of Bristol, Tyndall Avenue, Bristol BS8 1TL, UK}

\author{David Ernsting}
\affiliation{H.H. Wills Physics Laboratory, University of Bristol, Tyndall Avenue, Bristol BS8 1TL, UK}

\author{Wolfgang Kreuzpaintner}
\affiliation{Physik-Department, Technische Universit\"at M\"unchen, James-Franck Stra\ss e, 85748 Garching, Germany}

\author{Michael Leitner}
\affiliation{Heinz Maier-Leibnitz Zentrum (MLZ), Technische Universit\"at M\"unchen, Lichtenbergstr. 1, 85748 Garching, Germany}
\affiliation{Physik-Department, Technische Universit\"at M\"unchen, James-Franck Stra\ss e, 85748 Garching, Germany}

\author{Christian Pfleiderer}
\affiliation{Physik-Department, Technische Universit\"at M\"unchen, James-Franck Stra\ss e, 85748 Garching, Germany}

\author{Christoph Hugenschmidt}
\affiliation{Heinz Maier-Leibnitz Zentrum (MLZ), Technische Universit\"at M\"unchen, Lichtenbergstr. 1, 85748 Garching, Germany}
\affiliation{Physik-Department, Technische Universit\"at M\"unchen, James-Franck Stra\ss e, 85748 Garching, Germany}

\date{\today}

\begin{abstract}

We determined the bulk electronic structure in the prototypical Heusler compound Cu$_2$MnAl by measuring the Angular Correlation of Annihilation Radiation (2D-ACAR) using spin-polarized positrons.
To this end, a new algorithm for reconstructing 3D densities from projections is introduced that allows us to corroborate the excellent agreement between our electronic structure calculations and the experimental data.
The contribution of each individual Fermi surface sheet to the magnetization was identified, and summed to a total spin magnetic
moment of $3.6\,\pm\,0.5\,\mu_B/\mathrm{f.u.}$. 

\end{abstract}

\maketitle{}

Heusler alloys exhibit a most diverse range of phenomena~\cite{Graf20111}. Amongst these are e.g. half-metallicity which was first predicted in Heusler systems in the early 80s \cite{Kuebler1983,Groot1983}, the most promising magnetic shape memory material Ni$_2$MnGa \cite{Ullakko1996}, and the zero gap semiconductor Fe$_2$TiSn~\cite{Slebarski2000}. 
Just recently, it was reported that ${\mathrm{Mn}}_{2}\mathrm{CoAl}$ represents a new class of materials, so called spin gapless semiconductors \cite{Ouardi2013}.
The nature of the electronic interactions in Heusler compounds is known to be rather delicate. Detailed knowledge of the electronic structure is vital for tailoring specific physical properties such as magnetism and electron spin-polarization, since features in the band structure depend very sensitively on the composition \cite{Kuebler1983,Sasioglu2008}.

Being the prototype of all Heusler alloys \cite{Heusler1903}, Cu$_2$MnAl has also become a model system for understanding the electronic correlations in this class of materials \cite{Kuebler1983,Krumme2011,Sasioglu2008,Galanakis2012}. In particular, the interplay between the localized $d$-electrons and the delocalized electrons in Mn based Heusler systems is still under discussion, and in this context the shape of the Fermi surface is a key ingredient of the Ruderman-Kittel-Kasuya-Yosida (RKKY) interaction \cite{Sasioglu2008,Ruderman1954,Kasuya1956,Yosida1957}. Besides this fundamental questions, Cu$_2$MnAl has also a large relevance for applied physics as neutron polarizer/monochromator material \cite{Delapalme1971,Neubauer2012}.

A powerful experimental technique which can provide unique information about the bulk electronic structure is the measurement of the Two Dimensional Angular Correlation of electron-positron Annihilation Radiation (2D-ACAR) \cite{Dugdale2014a}.\ \textsl{Spin-polarized} 2D-ACAR has been used to prove half-metallicity in NiMnSb \cite{Hanssen1986,Hanssen1990} and to determine the electron-electron interaction strength in Ni \cite{Ceeh2014a}. Compared to Angle-Resolved Photo-Emission Spectroscopy (ARPES) the (high energy) positron probing the bulk is not affected by the surface, and the photon-matter interaction (which can complicate the analysis of ARPES data) does not have to be considered.

In this Letter we report spin-resolved 2D-ACAR measurements on a full Heusler compound in order to reveal the spin-polarized Fermi surface which is thought to play an important role in mediating the magnetic interactions.
We show experimentally how the contributions from majority and minority bands can be separated yielding the effective magnetic moments of each Fermi surface sheet.
Using our novel algorithm we reconstructed the spin-polarized 3D electron-positron momentum density (often referred to as the Two-Photon Momentum Density (TPMD) or $\rho^{2\gamma}$). Furthermore, we scrutinize the experimental results against band-theoretical calculations.

\begin{figure*}
	\includegraphics{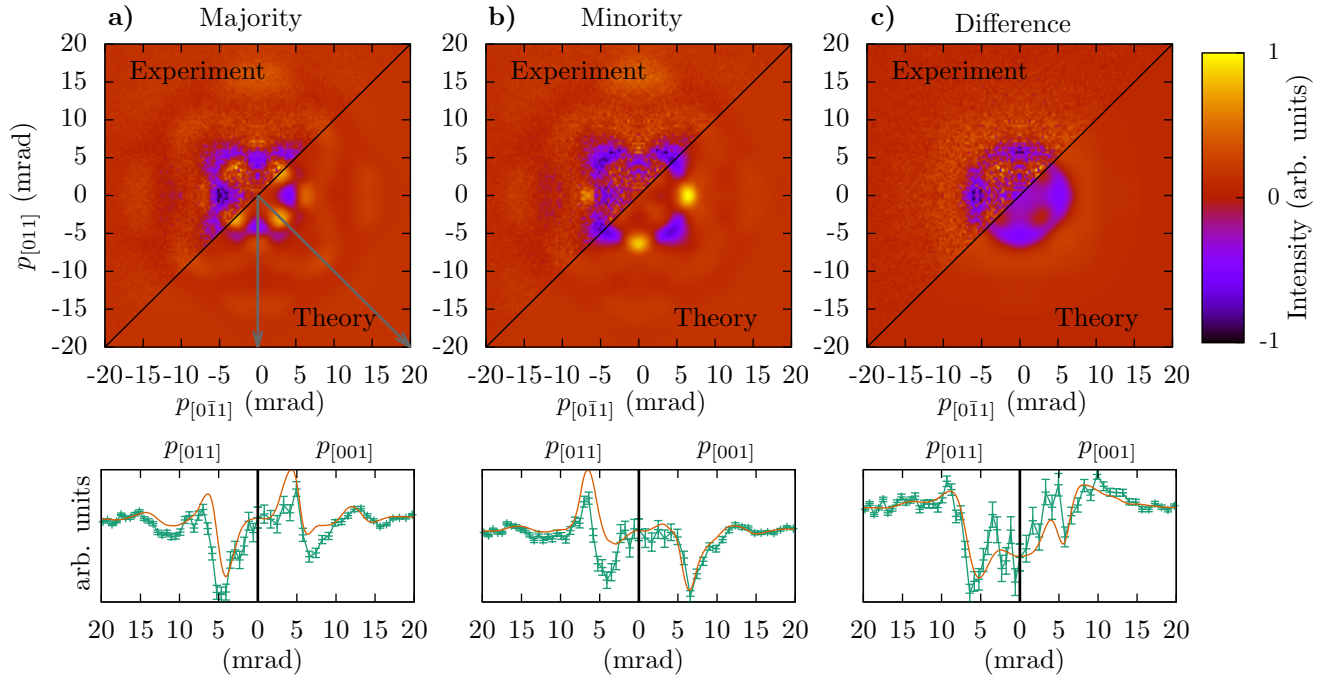}
	\caption{2D projections along $[100]$ of Cu$_{2}$MnAl at room temperature.
	The measured anisotropy of (a) majority and the (b) minority density respectively is compared with the theoretical results. The theoretical spectra were convolved with the resolution function of the spectrometer. The column (c) depicts the difference between majority and minority density with the total number of counts in each spectrum normalized to one. The lower row shows cuts through the upper pictures as indicated by the arrows in a) along  $[001]$ and $[011]$ directions for theory (orange) and experiment (green).}

	\label{fig:vglTheo}
\end{figure*}

When a positron is implanted in a solid, it will thermalize on a timescale of a few picoseconds and subsequently annihilate with an electron, conserving energy and momentum and predominantly producing two $\gamma$-photons. The transverse momentum in particular is conserved by an angular deviation from collinearity. This deviation is measured by coincidence detection of the $\gamma$-photons with position resolving detectors, yielding a 2D projection of the momentum of the annihilated pair \cite{Dugdale2014a}. Importantly, the momentum of the pair is dominated by that of the electron, with the finite positron momentum broadening the resolution.
Thus a projection $M$ of $\rho^{2\gamma}$ along the direction $p$ can be measured:
\begin{equation}
M(p_{\perp1},p_{\perp2})=\int_{-\infty}^{\infty}  \mathrm{d}p \; \rho^{2\gamma}(\mathbf{p}),
\end{equation}
where  $p$, $p_{\perp1}$ and $p_{\perp2}$ are three orthogonal directions in momentum space.

$\rho^{2\gamma}$ can be approximated as the sum (over all occupied electron states, $i$, and bands, $j$) of the Fourier transform of the product of positron $\Psi^{+}$ and electron $\Psi^{-}$ wavefunction \cite{Jarlborg1987}:
\begin{equation}
	\rho^{2\gamma}(\mathbf{p})\propto\sum_{\mathrm{occ.} i}\sum_j \left| \int \mathrm{d}\mathbf{r} \; e^{-i\mathbf{pr}}\Psi^{-}_{i,j}(\mathbf{r})\Psi^{+}(\mathbf{r}) \; \sqrt{\gamma(\mathbf{r})} \right|^{2},
\end{equation}

where $\gamma(\mathbf{r})$ is called the enhancement factor which takes into account the electron positron correlations. In the Independent Particle Model (IPM), $\gamma=1$ but often better descriptions are needed for quantitative agreement \cite{Laverock2010}.

Owing to parity violation in the weak interaction, positrons created in $\beta^{+}$-decay are longitudinally spin-polarized \cite{Zitzewitz1979}. The fraction of polarized positrons, $P$, is defined by $P={N^{\uparrow}}/{\left(N^{\downarrow}+N^{\uparrow}\right)}$, with the numbers of positrons with their spins parallel and antiparallel to the direction of magnetization of the sample being $N^\uparrow$ and $N^\downarrow$, respectively.
The mean emission energy of \Na{} yields a maximum $P=0.835$, although in practice it is reduced by backscattering inside the source and
by geometrical factors. In our experimental configuration it has been measured to be  $P=0.673$ \cite{Ceeh2014a}.

In a ferromagnet, the lifting of the degeneracy of the energies of electrons with opposite spins leads to there being more electrons of one spin (majority electrons) than the other (minority electrons).
Since in metals the electron-positron pair annihilates overwhelmingly in the spin singlet configuration, positrons with different polarizations will annihilate predominantly with electrons from either the majority or the minority spin channel. By reversing the polarity of the magnetic field at the sample position, the magnetization of the sample can be reversed. Making the reasonable assumptions that the sample is fully magnetized and the 3$\gamma$ annihilation can be neglected, we can express our measurement as a linear combination of the electron positron momentum density of the majority spin channel ($\rho_\mathrm{maj}$) and the minority spin channel ($\rho_\mathrm{min}$):
\begin{equation}
M^{p/a}\propto P\frac{\rho_\mathrm{maj/min}}{\lambda_{\mathrm{maj/min}}}+(1-P)\frac{\rho_\mathrm{min/maj}}{\lambda_{\mathrm{min/maj}}}
\label{eq:messp}
\end{equation}

for a magnetic field pointing parallel ($p$) and antiparallel ($a$) to the direction of positron emission, where $\lambda$ represents the annihilation rates for the majority and minority spin channels, respectively.

If both $M^p$ and $M^a$ are measured, straightforward algebra shows that it is possible to isolate the majority and minority spectra. Further insight into spin-polarized 2D-ACAR measurements can be found elsewhere \cite{Berko1964,Hanssen1990,Ceeh2014a}.

A single crystal disc-shaped sample of Cu$_2$MnAl with a diameter of 8\,mm and a height of 1\,mm was prepared and oriented with a $(011)$ face by x-ray Laue back-reflection, and its surfaces polished~\cite{Neubauer2012}.
The spin magnetic moment was determined via Compton scattering at \SI{300}{\kelvin} and a field  of \SI{1}{\tesla} to be $3.2\,
\mu_{\mathrm{B}}/\mathrm{f.u.}$ \cite{Duffy2015}. This is in good agreement with previously published values \cite{Felcher1963,Endo1964,Michelutti1978,Takata1965}.

Complementary positron annihilation experiments performed on our sample revealed that the vacancy density is below \num{9.7e-5} per atom~\cite{Hugenschmidt2015}.

The measurements were carried out at the 2D-ACAR spectrometer at the Technische Universit\"at M\"unchen \cite{Ceeh2013}.
We recorded spectra at five different projection angles in the $(011)$ plane, namely along
$[0\bar{1}1]$ and $[100]$ and three further projections at angles  \SI{29.8}{\degree}, \SI{35.3}{\degree}, and \SI{59.8}{\degree} with respect to the $[100]$ direction at room temperature. At each angle we took data for opposite sample magnetizations in a field of \SI{1.0}{\tesla}, collecting typically \num{1.3e8} counts.

With 2D-ACAR, 2D \textit{projections} of $\rho^{2\gamma}$ are measured. Nevertheless, the 3D $\rho^{2\gamma}$ can be recovered by measuring a series of 2D projections at different angles. Several methods have been applied to solve this inverse problem, and can be grouped in three categories, namely direct transform methods, methods using function expansion and iterative methods \cite{KontrymSznajd2009}. 
To use an iterative method it is necessary to express the measurement as a linear operator, $T$. To find a reconstruction $\mathbf{x}$, we can construct a least-squares-based function $f(\mathbf{x})$, which has to be minimized:
\begin{equation}
f(\mathbf{x})=\sum_{\alpha,i}\frac{\left(M_{\alpha}-T_{\alpha}\mathbf{x}\right)_{i}^{2}}{\sigma_{\alpha,i}^{2}}
\label{eq:fvonx}
\end{equation}
where $\alpha$ indexes the measured projections, and $i$ represents the data points of the spectrum. The estimated error of the measurement is expressed by $\sigma$. Since this minimization problem is underdetermined with the typical number of measured projections, it is necessary to apply a regularization functional. The most popular choice for the reconstruction of 2D-ACAR data, and the one used here, is an entropy-like function $g(\mathbf{x})=\sum_{i}x_{i}\ln(x_{i})$ \cite{Pylak2011}, although other regularization functionals have been investigated, too \cite{Weber2014,Leitner2015}. 

The maximum potential of iterative methods has not yet been exploited for experimental 2D-ACAR reconstructions. Here we show its power by means of an algorithm that uses the full crystal symmetry, properly accounts for the experimental resolution (previous approaches have deconvoluted the resolution separately \cite{Fretwell1995}), and preserves the statistical errors of the data through correction with the Momentum Sampling Function (MSF). The MSF is a geometrical correction of the data because of the finite size of detectors~\cite{West1995}. 
A similar approach with a parametrized Fermi surface has been proposed by Leitner {\it et al.}~\cite{Leitner2015}. Here we have implemented the operator $T_{\alpha}$ in Eq.~\ref{eq:fvonx} to comprise a projection operator $R_{\alpha}$, creating a projection from the density $\mathbf{x}$ in the irreducible wedge onto a plane, a convolution operator $C$ with the resolution function of the experimental setup, and a diagonal matrix $S$ with the values of the MSF:
\begin{equation}
T_{\alpha}=R_{\alpha} C S n_{\alpha},
\end{equation}
where the scalar factor $n_{\alpha}$ corresponds to the number of counts in a measurement $n_{\alpha}=\sum_{i} M_{\alpha,i}$. Thus, the reconstruction $\mathbf{x}$ will be normalized and the tensor $T_{\alpha}$ can be interpreted as the simulated measurement of a test density $\mathbf{x}$.

To complement the experimental results, the electronic structure was calculated using the ELK APW+lo code \cite{ELK} with the by x-ray diffraction experimental determined lattice constant of our sample ($a=5.961$ \AA).  
The generalized gradient approximation (GGA) was used to approximate the exchange-correlation functional \cite{Perdew1996}, and the resulting band structure was found to be in reasonable agreement with previous calculations \cite{Ishida1978,Deb2000}. The calculated spin moment was 3.51~$\mu_B/\mathrm{f.u.}$. In order to simulate the slightly reduced spin magnetic moment at room temperature \cite{Neubauer2012}, calculations were also performed at a fixed spin moment of 3.2~$\mu_B/\mathrm{f.u.}$.

Since the positron is delocalized, its density is vanishingly small across the macroscopic sample. This means that the positron wavefunction can be calculated from the self-consistent electron Coulomb potential (with opposite sign) together with an electron-positron correlation potential.
Here, the parameter-free GGA extension of the electron-positron correlation potential and enhancement factors parametrized by Drummond {\it et al.}~(hereinafter referred to as ‘DR’ enhancement) are used ~\cite{Drummond2011,Barbiellini2015}.
The momentum distribution was calculated from the wavefunctions  using a tetrahedron interpolation method \cite{Ernsting2014}.

First we show that our measurement could separate contributions from the majority and minority spin electrons from the spectra with opposite magnetic field directions, using the solution of Eq.\,\ref{eq:messp}. The anisotropy of these distributions is obtained by subtracting the radial average of the spectrum. In the left upper halves in Fig.~\ref{fig:vglTheo}\,a) and b) the anisotropy is shown for the $[100]$ projection.
The isotropic part in a spectrum has two main origins, namely annihilation with tightly bound core electrons and the annihilation in defects \cite{Dugdale2014b}. The pronounced anisotropic signal, persisting to quite large momentum, is a strong indication of the high quality of the crystal.
It is also clearly revealed that the majority and minority densities have strikingly different anisotropies, which in this case is mainly due to the two spin channels having different Fermi surface topologies (since most of the fully occupied bands are common to both spins).

In the right hand lower halves of Fig.~\ref{fig:vglTheo} we present the results of the electronic structure calculations.  These theoretical spectra were convolved with the point spread function of the spectrometer. The agreement between theory and experiment is excellent. The small deviation from four-fold symmetry in the $[100]$  projection is due to the asymmetrical resolution function, which is qualitatively similar for theory and experiment. For example in the anisotropy of minority spin channel (Fig.~\ref{fig:vglTheo} b)) at approximately $(\pm$\SI{5}{\milli \radian},\SI{0}{\milli\radian}) a pronounced feature above the radial average can be seen, which is slightly broader than at the equivalent points  (\SI{0}{\milli\radian},$\pm$\SI{5}{\milli \radian}).

Fig.~\ref{fig:vglTheo}\,c) depicts the difference of the majority and minority density of the measured and calculated spectra, respectively. For this distribution, each spectrum was normalized to unity. Since no further information is needed to calculate the difference spectrum (note, in particular, that the polarization of the beam and the different annihilation rates do not influence the result), it is ideally suited to compare theory and experiment. The structure in this image is mainly due to the contribution of unpaired $3d$ electrons.

In order to assess the effect of positron enhancement we calculated the reduced $\chi^2$ of a fit to the experimental data using the IPM and DR enhancement, respectively.

When enhancement is included in the theory, its agreement with experiment improves significantly for the sum (\num{4.4e2} to \num{2.0e2} for the $[100]$ and \num{5.3e2} to \num{2.4e2} for the $[0\bar{1}1]$ projection), as expected. 
In contrast, the enhancement only marginally improves the fit for the difference (\num{1.176} to \num{1.173} for the $[100]$ and \num{1.190} to \num{1.187} for the $[0\bar11]$ projection), as some of the effects of the positron cancel out. 
The variation appears because the dominant signal in the difference spectra originates from the more localized Mn $3d$ electrons, where enhancement effects are smaller. 
It has to be emphasized, that this result confirms experimentally that the enhancement effects in magnetic difference measurements are reduced as theoretically described in references \cite{Biasini2006} and \cite{Rusz2007}.

Theoretical calculations for the energy bands and Fermi surfaces of Cu$_2$MnAl were performed by Ishida {\it et al.}~\cite{Ishida1978,Ishida1981}. They predicted that all the majority sheets are $\Gamma$-centered and hole-like, but the minority bands generate very small hole pockets and some larger electron-like pockets centered on the $X$-points. With an unrestricted spin moment (i.e. 3.51\,$\mu_{B}/\mathrm{f.u.}$) we can qualitatively reproduce the results of Ishida {\it et al.}. The two larger majority sheets are predicted to have very similar size and shape, and they intersect the Brillouin Zone (BZ) with necks at the $L$-points. The smaller majority hole sheet resembles an octahedron. All three majority sheets nearly touch each other between the $\Gamma$-point and the $X$-point.  However, if we fix the magnetic moment to 3.2\,$\mu_{B}/\mathrm{f.u.}$ as suggested by the experiment, the small minority hole pockets are not present anymore.

The $\rho^{2\gamma}$ for both majority and minority electrons were reconstructed as described earlier. According to the Lock-Crisp-West (LCW) theorem \cite{Lock1973}, the densities were folded back into the first BZ by a transformation from momentum space to crystal momentum.  Hence, the LCW transformation restores translational invariance and reinforces the discontinuities due to the Fermi surface, making it easier to see.

\begin{figure}
	\includegraphics{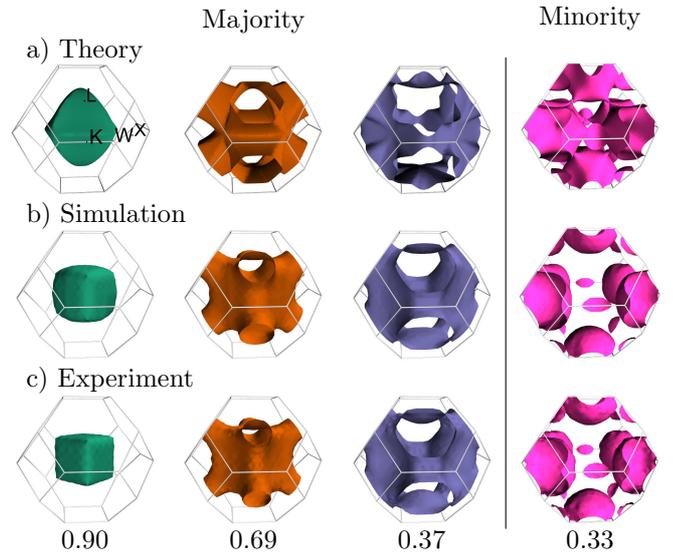}
	\caption{Majority (left) and minority sheets (right) of the Fermi surface of Cu$_2$MnAl. Spin-resolved Fermi sheets are obtained \emph{a)} from fixed spin moment (3.2~$\mu_B/\mathrm{f.u.}$) calculations, \emph{b)} from isosurfaces of reconstructed simulated data convolved with the experimental resolution function including statistical noise, and \emph{c)} from the isosurfaces of experimental data. The occupied fraction of the BZ volume of the experimentally determined Fermi surface is given below each sheet.}
	\label{fig:einzeln}
\end{figure}

The spin-resolved Fermi surface sheets of Cu$_2$MnAl from our fixed spin moment (3.2~$\mu_B/\mathrm{f.u.}$) theoretical calculation are presented in Fig.~\ref{fig:einzeln}~a). Additionally, Fig.~\ref{fig:einzeln}~b) shows a reconstruction using our iterative approach  of the calculated data taking into account statistical noise and the experimental resolution. The reconstruction from the experimental data is shown in Fig.~\ref{fig:einzeln}~c). Obviously, the Fermi surface sheets of experiment and simulation are in excellent agreement. As expected, the sharp Fermi-breaks in the 3D density become smeared out when the spectra are convolved and when noise is introduced. Nevertheless, we are able to determine the topology correctly and, moreover, can discern distinct features of the Fermi surface sheets, e.g. the small pockets at the K-point of the minority surface.

In Fig.~\ref{fig:einzeln}~c) the occupied fraction of the BZ is given for the experimentally determined Fermi surface sheets. We obtained a value for the total magnetization of $3.6\,\pm\,0.5\,\mu_B/\mathrm{f.u.}$, in good agreement with other magnetization measurements \cite{Duffy2015,Felcher1963,Endo1964,Michelutti1978,Takata1965}, if we assume that there are two additional completely filled bands in the majority spin channel (as also indicated in our calculations). Hence, approximately two thirds of the magnetic moment is contributed from tightly bound states, while one third of the magnetic moment stems from conduction electrons which are less tightly bound. However, even these electrons are not completely delocalized as \.{Z}ukowski {\it et al.}~reported~\cite{Zukowski1997}.

In conclusion, through spin-polarized 2D-ACAR measurements on the Heusler system Cu$_2$MnAl, we have demonstrated a novel approach for extracting spin-resolved Fermi surfaces. 
As predicted by theory, there are unoccupied states in all bands at the $\Gamma$-point. 

The experimentally determined Fermi surface sheets are shown to be in excellent agreement with the theory.  We emphasize that spin-polarized 2D-ACAR is a unique technique which offers a great potential for spin-resolved measurements of the bulk electronic structure in correlated systems at finite temperatures.

\begin{acknowledgments}
This project is funded by the Deutsche Forschungsgemeinschaft (DFG) within the Transregional Collaborative Research Center TRR 80 ``From electronic correlations to functionality''. 
\end{acknowledgments}

\bibliography{lit}
\end{document}